\documentstyle[11pt,paspconf,epsf]{article}

\def\half{{\scriptstyle {1 \over 2}}}

\def\onetwelfth{{\scriptstyle {1 \over 12}}}

\def\threehalf{{\scriptstyle {3 \over 2}}}

\def\dt{{\delta t}}
\def\ie{{\it {\frenchspacing i.{\thinspace}e. }}}

\def\et{{\it et~al.} }

\def\simlt{\hbox{ \rlap{\raise 0.425ex\hbox{$<$}}\lower 0.65ex\hbox{$\sim$} }}
\def\ltorder{\hbox{ \rlap{\raise 0.425ex\hbox{$<$}}\lower 0.65ex\hbox{$\sim$} }}
\def\simgt{\hbox{ \rlap{\raise 0.425ex\hbox{$>$}}\lower 0.65ex\hbox{$\sim$} }}
\def\gtorder{\hbox{ \rlap{\raise 0.425ex\hbox{$>$}}\lower 0.65ex\hbox{$\sim$} }}
\def\sles{\lower2pt\hbox{$\buildrel {\scriptstyle <}
   \over {\scriptstyle\sim}$}}
\def\sgreat{\lower2pt\hbox{$\buildrel {\scriptstyle >}
   \over {\scriptstyle\sim}$}}

\def\qquad{\quad\quad}

\newcommand{\be}{\begin{eqnarray}}
\newcommand{\ee}{\end{eqnarray}}

\begin{document}

\title{Time Symmetrization Meta-Algorithms}

\author{Piet Hut\altaffilmark{1},
        Yoko Funato\altaffilmark{2},
        Eiichiro Kokubo\altaffilmark{2},
        Junichiro Makino\altaffilmark{3},
        Steve McMillan\altaffilmark{4}
       }

\altaffiltext{1}{Institute for Advanced Study, Princeton, NJ 08540, U.S.A.}
\altaffiltext{2}{Department of Earth Science and Astronomy,
                 College of Arts and Sciences, University of Tokyo,
                 3-8-1 Komaba, Meguro-ku, Tokyo 153, Japan}
\altaffiltext{3}{Department of General Systems Study,
                 College of Arts and Sciences, University of Tokyo,
                 3-8-1 Komaba, Meguro-ku, Tokyo 153, Japan}
\altaffiltext{4}{Department of Physics and Atmospheric Science,
                 Drexel University, Philadelphia, PA 19104, U.S.A.}

\begin{abstract}
We present two types of meta-algorithm that can greatly improve the
accuracy of existing algorithms for integrating the equations of
motion of dynamical systems.  The first meta-algorithm takes an
integrator that is time-symmetric only for constant time steps, and
ensures time-symmetry even in the case of varying time steps.  The
second meta-algorithm can be applied to any self-starting integration
scheme to create time-symmetry, both for constant and for variable
time steps, even if the original scheme was not time-symmetric.  Our
meta-algorithms are most effective for Hamilton systems or systems
with periodic solutions.  If the system is not Hamiltonian (for
example, if some dissipative force exists), our methods are still
useful so long as the dissipation is small.
\end{abstract}

\section{Introduction}

Many problems in computational physics are governed by underlying
equations that are intrinsically time-symmetric.  In particular, for
any simulation in Hamiltonian dynamics, we can run the movie of our
computation equally well forwards as backwards, and in both cases
obtain a physically allowable solution.  Clearly, it is desirable to
use an integration algorithm that reflects this time-symmetry as a
built-in property.  In that case, intuitively speaking, particles can
no longer fly `off the tracks', so to speak, when moving through a
curve.  An example of particular interest to this conference is
astrophysical particle simulations, in which particles may correspond
directly to physical units, such as molecules or stars, or may form
tracers used to approximate the solution of a set of underlying
equations with continuous variables.

The basic idea is this: if an algorithm would make a particle spiral
out systematically in a given situation, it would have to do so
equally in the forward and backward direction.  Time reversal,
however, would force an inward motion, and the conclusion is that the
net spiral out (or spiral in) has to be zero, at least when averaged
over a full period in a periodic system.  The error in energy thus has
to be periodic as well, and cannot build up from orbit to orbit.  Of
course, errors are still made for finite integration step size, but
they show up as errors in phase.  In many applications, phase errors
are preferable over errors in quantities such as the total energy,
that should be conserved.

Most algorithms that are in common use for orbit integration in
simulations are not time-symmetric.  Even those that are, typically
lose their symmetry property as soon as one allows variable time steps.
In the remainder of this paper, we offer two types of meta-algorithm
for constructing a larger class of time-symmetric algorithms.  In
\S\S2,3 we show how to restore time symmetry for algorithms that were
symmetric, but lose that property when one uses variable time steps.
In \S\S4,5 we show how to create time symmetry even for those
algorithms that were never time symmetric to begin with, not even in
the constant time step case.

\section{Building A Better Leapfrog}

A celebrated example of an integration scheme with built-in time
symmetry is the leapfrog scheme, also known as the Verlet method.  It
is widely used in many applications of particle simulations, such as
in molecular dynamics, plasma physics, fluid dynamics and stellar
dynamics(Hockney \& Eastwood 1988; Barnes \& Hut 1986).  The time
symmetry is manifest in the interleaved representation, which gave
rise to the `leapfrog' name:
$$
r_1 = r_0 + v_\half \dt,\eqno(1a)
$$
$$
v_\threehalf = v_\half + a_1\dt,\eqno(1b)
$$
where $r$ can stand for the position vector of a single particle or
the combined vector ${\bf r}_1, {\bf r}_2,$ $\dots,$ ${\bf r}_N$
representing a system of $N$ particles.  The quantity $v = dr/dt$ is
the velocity and $a(t) = a(r(t)) = dv/dt$ the acceleration.  The
subscripts after the various quantities indicate the time at which
they apply, in units of the time step, \ie $v_\half = v(t+\half
\dt)$.

It is convenient to map the standard interleaved description into a
form in which all variables are defined at the same instant in time:
$$
r_1 = r_0 + v_0\dt + \half a_0(\dt)^2,\eqno(2a)
$$
$$
v_1 = v_0 + \half(a_0 + a_1)\dt.\eqno(2b)
$$
Starting from $\{r_0,v_0,a_0\}$, one first computes $r_1$, then
$a_1(r_1)$, by evaluating the appropriate expression dictated by the
system under consideration, and finally $v_1$.  While Eq.(2) looks as
though it has lost its explicit time symmetry, it is still equivalent
to the original Eq.(1), as can be verified by direct substitution
(Barnes \& Hut 1989).  However, if the time step $\dt$ is allowed to
vary, through a functional dependence $\dt = h(r,v)$ for example, time
symmetry is lost.

Time symmetry can be restored if we force the time step to be a
symmetric function of the begin point and end point of each time step,
as was first shown by Hut \et (1995).  For example, we can use
$$
\dt = \half[h(r_0,v_0)+h(r_1,v_1)],\eqno(3)
$$
With this choice of time step, the combined set of equations (2,3) has
become implicit: in order to determine $\{r_1,v_1\}$ from
$\{r_0,v_0\}$, we need to know the time step size $\dt$, which in turn
is dependent on $\{r_1,v_1\}$.  We can solve this problem by starting
with $\dt = h(r_0,v_0)$ as a first approximation.  This will give us
approximate values for $\{r_1,v_1\}$, from which we can determine a
more accurate value for $\dt$ using equation (3).  If necessary, this
iteration process can be repeated several times, but in practice one
or two iterations are generally sufficient to reach time symmetry to
machine accuracy.

\section{Restoring Time Symmetry}

The notion of time symmetry restoration for variable time steps, give
above for the particular case of the leapfrog scheme, carries over to
any class of integration schemes that is explicitly time symmetric for
constant time steps, as shown by Hut \et (1995).
Another example, given in the same paper, concerns the following natural
fourth-order generalization of the leapfrog scheme:
$$
r_1 = r_0 + \half(v_1 + v_0)\dt - \onetwelfth(a_1-a_0)(\dt)^2,\eqno(4a)
$$
$$
v_1 = v_0 + \half(a_1 + a_0)\dt - \onetwelfth(j_1-j_0)(\dt)^2,\eqno(4b)
$$
which is a truncated form of the Hermite scheme (Makino 1991a).
Here the jerk $j = da/dt$ is calculated directly by differentiation
of the expression for the force (thereby introducing a dependency on
velocity as well as position in the case of Newtonian gravitational
forces).  This set of equations is manifestly time symmetric for
constant time steps.  Unlike the original second-order leapfrog, given
in eq. (2), the scheme in eq. (4) is implicit, already for constant time
steps.  Applying the same symmetrization procedure given in equation (3)
leads, after iteration, to a fully time-symmetric fourth-order
generalization of the leapfrog.  In practice, iteration can give a quick
convergence, leading to substantially improved accuracy for a fixed
amount of computer time (Hut \et 1995).

\section{Creating Time Symmetry}

Even if no time symmetry is present in a given algorithm, it is
possible to construct an implicit version that is manifestly time
symmetric, for the general class of self-starting (\ie one-step)
integration schemes.  The meta-algorithm that performs this feat is
described in detail by Funato \et (1997), as a generalization of
the prescription offered by Hut \et (1995).  Here we give a brief
outline of the main idea behind the treatment.

Let us start with the ordinary differential equation
\be
\frac{dy}{dx} & = & {f}(x,y),
\ee
and a given self-starting integration scheme, expressed as
\be 
y_{i+1} & = & y_{i} + F(x_{i},y_{i};h_i),
\ee
where $x_i$ and $y_i$ are the values of the variables $x$ and $y$
after the $i$-th step, and $h_i$ is the step size at $x_i$.

Our meta-algorithm can be expressed as follows:
\be
y_{i+1} & = & y_{i} + \tilde F(x_{i},y_{i};{\tilde h}_i) =
y_{i} + \frac{1}{2}\left[F(x_{i},y_{i};{\tilde h}_i) - 
F(x_{i+1},y_{i+1};-{\tilde h}_i) \right],
\label{eqn:meta}
\ee
where ${\tilde h}_i$ can be constructed in the form of a function
${\tilde h}_i= f(h_i, h_{i+1})$ that is symmetric in its arguments:
$f(x,y)=f(y,x)$.  For example, we could take simply
\be 
{\tilde h}_i = \half[h(x_i,y_i)+h(x_{i+1},y_{i+1})],
\ee
or a root mean square, or any other symmetric combination.
Equation (\ref{eqn:meta}) gives an implicit formula for $y_{i+1}$.  
As before, we can solve this equation by iteration, starting with
the original non-symmetric scheme as the initial trial function.

While this recipe is surprisingly simple, we can do even better.
Instead of taking a given estimate for the step size, we can use
the difference 
\be
\Delta F_{i} \equiv \tilde F(x_{i},y_{i},{\tilde h}_i) -
F(x_{i},y_{i},{\tilde h}_i), 
\ee
to estimate the local truncation error.  This information can be used
to implement a form of adaptive step size control.  See Funato \et
(1997) for further details.

\section{Building A Better Runge-Kutta}

\begin{figure}

\leavevmode
\epsfxsize 6.5cm
\epsffile{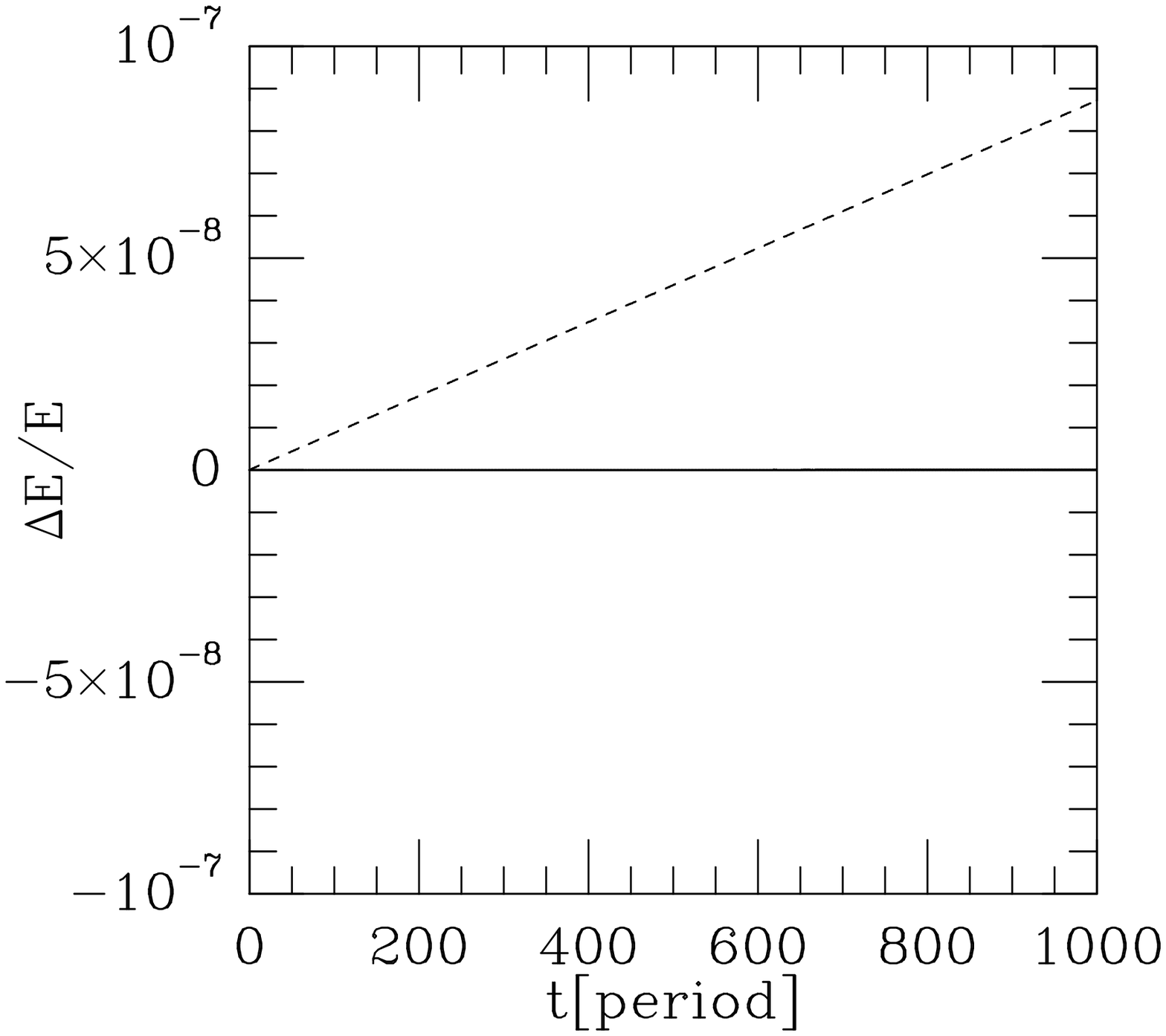}
\epsfxsize 6.5cm
\epsffile{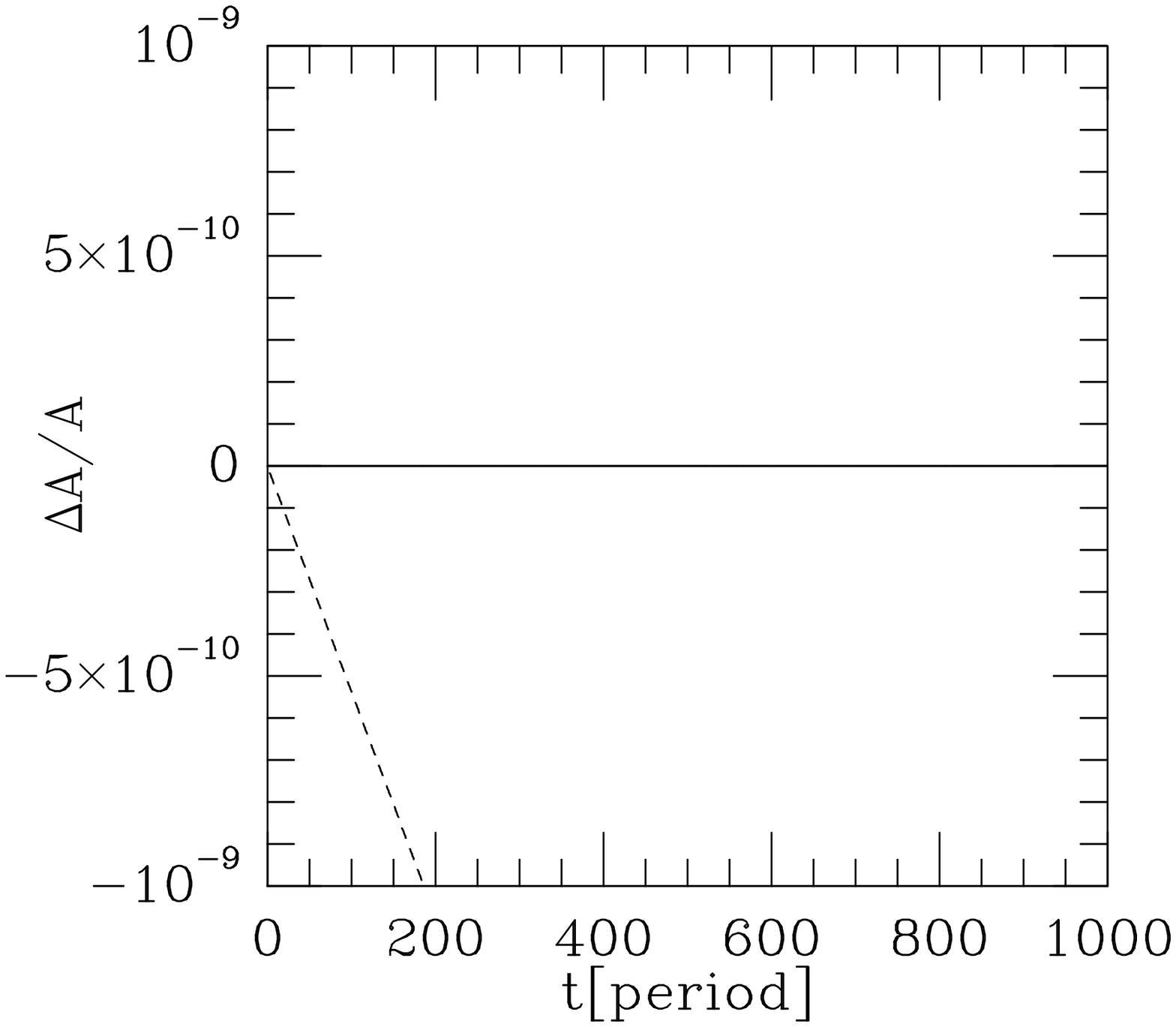}

\caption{Effects of our meta-algorithm applied to a fourth-order
Runge-Kutta scheme.  Plotted is the growth of the relative error in
energy (left) and angular momentum (right) for a Kepler orbit with
eccentricity 0.9.  Dashed and full lines correspond to the standard
Runge-Kutta scheme and the time-symmetrized Runge-Kutta version,
respectively.}
\label{fig1}
\end{figure}

As an example application of our more general meta-algorithm, we have
constructed a time-symmetric version of the popular fourth-order
Runge-Kutta integration scheme.  Figure \ref{fig1} shows the behavior
of the errors in the relative energy and angular momentum for a binary
orbit with initial eccentricity $e = 0.9$ (the values plotted are
determined at apocenter).  The dashed and solid curves show the time
evolution of the errors for the standard Runge-Kutta scheme and for
the symmetrized Runge-Kutta scheme, respectively.  In both cases,
variable step sizes have been used.  The number of time steps is
comparable in both cases (around 600 per orbit).

Figure \ref{fig1} shows that no discernible secular error is produced,
even after 1000 orbital periods, for the run integrated by the
time-symmetric fourth-order Runge Kutta Method.  In contrast, the
error increases linearly for the run integrated by the standard
fourth-order Runge Kutta Method.  Further quantitative details will be
provided in the cost/performance analysis by Funato \et (1997).

\section{Discussion}

We have reviewed two types of meta-algorithm, based upon a
time-sym\-me\-tri\-za\-tion procedure.  The first meta-algorithm preserves
time symmetry that would otherwise be lost when integration step
sizes are allowed to change during integration.  The second
meta-algorithm creates time symmetry, even for those algorithms where
no symmetry was present in the equal-step-size case.

We mention here briefly a few recent applications of these ideas.
McMillan \& Hut (1996) have constructed a fully automated package for
performing gravitational three-body scattering experiments, where the
central orbit integrator is built along the principles outlined by Hut
\et (1995).  They found that the symmetrization meta-algorithm gave a
significant speed-up to the fourth-order Hermite scheme used.  Most
importantly, they found that the fraction of rejected experiments was
diminished greatly, compared to the standard integration scheme.  The
problem here is that some resonant scattering experiments can stay in
an intermediate state for a very long time, before finally decaying
into a final state.  No matter how accurate an initial time step
criterion has been chosen, there will always been a small fraction of
such lingering states that will ultimately lead to an unacceptable
build-up of errors (both integration errors and round-off errors).
Time symmetrization, while not circumventing this problem, can greatly
alleviate the situation.

Another application has been discussed by Funato \et (1996a,b).  They
have applied time symmetrization to the Kustaanheimo-Stiefel
regularization method, a sophisticated way to `unfold' the singularity
of the three-dimensional Kepler problem by mapping each point in three
dimensional space to a unit circle in an auxiliary four dimensional
space.  The combination of these two powerful techniques has resulted
in the most accurate way yet designed to integrate orbits near
collision singularities.

For some applications, specific adaptations of our general
meta-algorithm can make good use of the known constraints inherent in
the underlying problem.  One example that we have recently explored is
that large-scale simulations of planetary formation.  The problem is
that close encounters and physical collisions of planetesimals make it
absolutely necessary to use individually variable time steps (Aarseth
1985).  All standard choices for highly accurate integration schemes,
such as the symplectic schemes, lose their desirable properties once
we allow individual particles to change their integration time step
length at will.  In contrast, our meta-algorithm shows a way out, as
demonstrated by Kokubo \& Makino (1997).  They made use of the fact
that most planetesimals have nearly circular orbits ($e<<0.01$), which
means that their time step is practically constant when block time step
(McMillan 1986, Makino 1991b) are used, even if we allow the use of
variable time steps.  Even though the time step size of a particle
shrinks significantly during a close encounter, leading to a
break-down of time-symmetry, this occurs only for a very small
fraction of time, for a typical particle.  Since most of the
integration error actually comes from the integration of nearly
unperturbed orbits around the sun, reserving strict time symmetry for
unperturbed orbits turns out to be a good compromise, leading to high
overall accuracy.  Details will be provided by Kokubo \& Makino
(1997).

\acknowledgments

This work was supported in part by the National Science Foundation
under grants ASC-9612029 and AST-9308005, and by a Grant-in-Aid for
Specially Promoted Research (04102002) of the Ministry of Education,
Science, Sports and Culture, Japan.

\end{document}